\documentclass[12pt,preprint]{aastex}

\shorttitle{The WR 38/WR 38a Cluster}
\shortauthors{Wallace et al.}

\received{2004 April 8}
\accepted{}

\begin{document}

\title{{Hubble Space Telescope Imaging of the WR 38/WR 38a Cluster}\altaffilmark{1}}

\author{Debra J. Wallace\altaffilmark{2}, Douglas R. Gies}
\affil{Center for High Angular Resolution Astronomy, 
Department of Physics and Astronomy, Georgia State University,
Atlanta, GA 30303}
\email{wallace@chara.gsu.edu, gies@chara.gsu.edu}

\author{Anthony F. J. Moffat}
\affil{D\'epartement de Physique and Observatoire du Mont M\'egantic, 
Universit\'e de Montr\'eal,  C.~P.~6128, Succ.\ 
Centre-Ville, Montr\'eal, Qu\'ebec, H3C 3J7, Canada}
\email {moffat@astro.umontreal.ca}

\author{Michael M. Shara}
\affil {Department of Astrophysics, American Museum of Natural
History, New York, NY 10024}
\email {mshara@amnh.org}

\and

\author{Virpi S. Niemela\altaffilmark{3}}
\affil{Facultad de Ciencias Astron\'omicas y Geof\'{\i}sicas, 
Universidad Nacional de La Plata, Paseo~del~Bosque s/n, 1900 La Plata, Argentina}
\email{virpi@lilen.fcaglp.unlp.edu.ar} 

\altaffiltext{1}{Based on observations with the NASA/ESA {\it Hubble Space Telescope}, 
obtained at the Space Telescope Science Institute, which is operated by AURA, Inc., 
under NASA contract NAS5-26555.}
\altaffiltext{2}{Present address: NASA/GSFC, Code 685, Building 21,
Greenbelt, MD, 20771, wallace@exo1.gsfc.nasa.gov}
\altaffiltext{3}{Member of Carrera del Investigador, CIC-BA, Argentina}

\begin{abstract}
We are conducting a high angular resolution imaging survey 
of Galactic Wolf-Rayet stars using the Wide Field and Planetary 
Camera 2 aboard the {\it Hubble Space Telescope}.  
We have found a small stellar cluster associated with 
the faint, close pair WR 38 and WR 38a.  
We present astrometric measurements and photometry in 
the wide-band F336W ($U$), F439W ($B$), and F555W ($V$) filter system 
for these cluster and nearby stars.   We combine our 
photometry with Johnson and IR magnitudes, and compare the 
observations with calibrated model results for reddened 
stars to adjust the HST zeropoints and to identify five probable main sequence members of the cluster.  
A least-squares fit of the colors and magnitudes of this set 
yields a cluster reddening of $E(B-V) = 1.63 \pm 0.05$ mag and 
a distance of $10^{+12}_{-4}$ kpc for an assumed ratio of total-to-selective
extinction of $R=3.1$.  We discuss the relationship of 
this cluster to other objects along the line of sight.  If situated at a distance 
of $\approx 8$~kpc, then the cluster would reside in a dense region of the Carina spiral arm, 
close to a giant molecular cloud and the starburst cluster NGC~3603. 

\end{abstract}

\keywords{binaries: visual --- stars: imaging --- stars: Wolf-Rayet
--- stars: individual (WR 38, WR 38a) ---
open clusters and associations: individual (C1104--610a)}

\section{Introduction} 

Wolf-Rayet (WR) stars are the evolved, He-burning cores and hot
stellar envelopes of once-massive O-type stars. Their strong winds
(terminal velocities: $v_{\infty}$~$\approx$~1000~-~2500~km~s$^{-1}$)
and heavy mass loss rates
($\dot{M}$~=~$\sim[1-6]~\times$~10$^{-5}$~$M_{\odot}$~yr$^{-1}$; 
\citealt{wil91}; \citealt{nug02}) power their
characteristic strong, broad emission lines of He, N, C, and O, 
which correspond to increasingly evolved evolutionary states.  While the
environment and multiplicity of O-type stars have been explored in
depth \citep{gie87,mas98}, these characteristics for
the evolved WR phase are not as well determined. In 1996 we initiated
a survey with the {\it Hubble Space Telescope} Wide Field and Planetary
Camera 2 ({\it HST} WFPC2) to discover and quantify the multiplicity and
environments of Galactic WR stars.  Here we discuss two of our
targets, WR~38 and WR~38a, that were revealed to be both multiple and
contained within a small stellar cluster \citep{wal99}.

WR~38 is a WC4-type star \citep{mac70} and WR~38a is a WN5-type
\citep{sha91}.  The two are separated by only $20\arcsec$ in the sky,
and they are neighbors to another even tighter $5\arcsec$ pair of WR
stars, WR~38b and WR~39, at a separation of $3\farcm8$.  These four
plus WR~37 are surrounded by a ring nebula \citep{mar97}, although it
is not clearly associated with any one of these stars.  Our view in
this direction in the Galactic plane cuts through the
Sagittarius-Carina spiral arm at distances of approximately 2.4 and
8~kpc \citep{gra88,geo00,rus03}.  The photometric distances of these
WR stars are generally close to 5~kpc \citep{van01}, which would place
them among the young population of the Carina spiral arm.  However, in
a recent paper \citet*{sho04} presented photometry for the stars in
the immediate vicinity of WR~38 and WR~38a that indicated a much
greater distance of $14.5 \pm 1.6$~kpc.  This distance would place the
pair in an extension of the outer Perseus arm some 14~kpc from the
Galactic center.  There are only 4 of 227 known Galactic WR stars that
have such a great distance from the Galactic center \citep{van01}.
\citet{sho04} designate the cluster as C1104--610a, and images of the
field surrounding the cluster are presented in \citet{wra76}
(Plate~4), \citet{van81}, and \citet{sho04} (Fig.~1 and 2).

Here we present new astrometric and photometric results on the stars
found close to WR~38 and WR~38a based upon {\it HST} PC images with
superb angular resolution.  We identify which stars are associated
with the WR pair based upon their derived reddening and location in a
color-magnitude diagram.  Our results are mainly consistent with those
of \citet{sho04} but we argue that the acceptable distance range is
such that the WR pair and its surrounding cluster may reside in the distant portion of the Carina spiral arm.

\section{Observations and Data Reduction} 

Observations of the fields of WR~38 and WR~38a were obtained in 1996 March
(JD 2450166.58) through the F336W, F439W, and F555W
filters as part of our {\it HST} WFPC2 survey of Galactic WR stars.
These WFPC2 filters were designed to correspond to the Johnson
$UBV$ filter set \citep{joh53,joh66}, 
but because of the strong emission lines present in WR spectra 
the transformation from the {\it HST} to the Johnson system is 
approximate at best for them.  Additionally, the WFPC2 F336W filter suffers from
a red leak which may cause dusty WR or other very red stars to appear 
brighter than in a similar $U$ band image. Therefore, all of the data
reductions and analyses were performed within the WFPC2 synthetic
magnitude system \citep{hol95}.

We centered the WR stars within the PC chip to take advantage of the
high resolution (scale of $0\farcs046$ pixel$^{-1}$).  
The observations also utilized a
high gain setting, with a threshold of $\sim$~53000 photoelectrons
pixel$^{-1}$, to optimize the discovery of faint companions. For each
observation we made two exposures to facilitate the removal of cosmic
rays (40~s for the F336W, 10~s for the F439W, and 2~s for the F555W filter).
Since these are SNAPSHOT-type observations, which are designed to fill
holes in the {\it HST} schedule, the observations were taken using only one
guide star. Fortunately, due to the short exposure times, this
resulted in no discernible loss of data quality due to telescope
motion, although this did place limits on the accuracy of the absolute
astrometry as the telescope may roll during the observation
\citep{bag02,gon02,nel02}.

WR~38 is sufficiently close to WR~38a that both stars could be
captured within the same WFPC2 image (Fig.~1).  The {\it HST}
image clearly shows a number of components surrounding both WR stars
that are not fully resolved in ground-based images 
\citep{sha91,sho04}.  Without spectroscopic
data we can only estimate the identity of the WR star within each
stellar grouping by assuming it is the brightest object within
the existing astrometric error box.  The numbering 
system we adopted is based upon angular separation from 
WR~38 (star \#1), and WR~38a is labeled as \#21.   

\placefigure{fig1}  

After normal pipeline processing 
\citep{bag02}, we aligned the six raw images using the task
IMALIGN in IRAF\footnote{IRAF is distributed by the National Optical
Astronomy Observatory, which is operated by the Association of
Universities for Research in Astronomy, Inc. under cooperative
agreement with the National Science Foundation.}/STSDAS. We then
combined each pair of images per filter using the routine CRREJ. 
We identified the stars in each combined image using the task DAOFIND
with the detection threshold set at $10\sigma$ over the background
level.  We determined a background value by using the IMEXAMINE
routine to derive the mean background value within $5\times5$ 
pixel boxes located at 5--10 well-distributed positions across the image. 
The single average of these values was used for the detection threshold.

Because the observations were made with just one guide star, 
the images may suffer from telescope roll and jitter. 
Consequently, the point spread function (PSF) of the stellar images 
may be asymmetrical, leading the DAOFIND algorithm to select false stars.  
Thus, all the detected stars were visually inspected before acceptance.

We used this list as input to the IRAF/STSDAS package METRIC to derive
the astrometric information given in Table~1.  METRIC
translates the WFPC2 pixel positions to celestial coordinates 
after correcting for geometric distortion.  The relative
positions of well-exposed stars are accurate to better than
$0\farcs005$ for targets imaged on one chip (errors are primarily
due to centering uncertainties).  However, while the relative positions
are highly accurate, the absolute positions of the objects are subject
to larger errors of $\pm(0.5-5)$ arcseconds due to the use
of a single guide star \citep{bag02,gon02,nel02}.
Fifteen of the stars in Table~1 appear in the 
{\it 2MASS All-Sky Catalog of Point Sources} \citep{cut03},
and the mean differences in the coordinates are 
$\alpha({\rm 2MASS}) - \alpha({\rm HST}) = +0.175 \pm 0.008$~s and
$\delta({\rm 2MASS}) - \delta({\rm HST}) = +0\farcs14 \pm 0\farcs18$.

\placetable{tab1}      

We performed aperture photometry using the IRAF package DAOPHOT \citep{ste87}. 
We chose aperture photometry over PSF photometry because of the
difficulties involved in fitting the PSF due to the small telescope
motions.  We used an 11 pixel ($0\farcs5$) radius aperture to 
determine the total stellar flux, and we estimated the sky background
using the OFILTER algorithm within the PHOT package of
DAOPHOT (as recommended by \citealt{fer96}). 

A number of photometric corrections are required to transform from raw
to calibrated magnitudes \citep{bag02}. The stellar
magnitudes were corrected for charge transfer efficiency using the
\citet*{whi99} formula for faint sources and the
\citet{dol00}\footnote{\url{http://www.noao.edu/staff/dolphin/wfpc2\_calib/}}
formula for more luminous sources (brighter than $100
e^{-}$).  Geometric distortion effects were corrected using a scheme that
rescaled the flux measurements of the stars based on their positions
within the image.  Contamination corrections were made using
information taken from the WFPC2 Web
site\footnote{\url{http://www.stsci.edu/instruments/wfpc2/}}. 
The $34^{th}$ row effect was corrected using the scheme of
\citet{and99}.  Most significantly, the zero-point corrections
for the F555W and F439W magnitude data were calculated using the 
\citet{dol00} updates to the original \citet{hol95} offsets.  
The \citet{dol00} revisions to the \citet{hol95} zero-points ($\approx 0.01$ mag) 
are not the result of temporal effects, but rather of improved calibration
methods, and hence, they have greater accuracy.  Unfortunately, Dolphin
did not provide updates for the F336W filter, so we used the 
\citet{hol95} calibration for the F336W filter images.

Our final magnitudes in the {\it HST} system are collected in Table~2. 
We have included a transformation of the {\it HST} wide-band magnitudes 
to Johnson $BV$ using the scheme of \citet{dol00}. 
We caution that this scheme may not be reliable for the 
WR stars since both the $B$ and $V$ band-passes are sensitive to 
the strong emission lines present in WR spectra \citep*{van87}. 
We did not attempt to transform our F336W magnitudes to Johnson $U$ 
since the latter straddles the Balmer jump while the former 
is sensitive mainly to shorter wavelength flux. 
We have also included in Table~2 the corresponding identification number (STP) and magnitude information from the work of \citet{sho04}
for 7 stars in common (mainly cluster members). 

\placetable{tab2}      

We found that there are systematic differences between our transformed 
$V$ and $B-V$ magnitudes compared to those of \citet{sho04}. 
We omitted the WR stars 
from this comparison because of our concerns about the transformation 
to Johnson magnitudes, and this leaves 5 stars in common between 
our samples.  We find mean differences of 
$<V_{\rm HST} - V_{\rm STP}> = 0.24 \pm 0.03$ and 
$<(B-V)_{\rm HST} - (B-V)_{\rm STP}> = -0.12 \pm 0.09$, 
i.e., the mean differences are larger than the 
standard deviations from the mean differences (and the standard 
deviations are comparable to the expected observational errors).  
These differences are probably due to errors in the adopted 
zero-points for the WFPC2 images.  We have found, for example, 
that there is an intrinsic scatter of order 0.1~mag between the 
observed and synthetic {\it HST} magnitudes for some two dozen WR stars with 
NUV spectroscopy from the {\it International Ultraviolet Explorer Satellite} 
and optical spectrophotometry \citep{tor87,tor88}.  
While part of this scatter may be due to errors in the 
calibration of the spectrophotometry, we caution that zero-point 
offsets of order 0.1~mag may be required to place our photometric 
results on an absolute scale.  Thus, the relative magnitudes for stars 
within a given image are secure, but color indices formed from magnitude 
differences between images may require a zero-point shift. 
In the next section we describe one way of estimating the 
zero-point adjustments based upon a reddening analysis of 
multi-wavelength photometry. 

\section{Interstellar Reddening}  

Extinction due to interstellar dust decreases at longer wavelengths, 
so the best approach to determining reddening and extinction is 
to compare both long and short wavelength flux measurements with 
spectral models transformed using a suitable reddening law. 
We were pleased to learn that 15 of the 25 targets in our 
{\it HST} field are included in the 
{\it 2MASS All-Sky Catalog of Point Sources} \citep{cut03} 
that lists $J$, $H$, and $K$ magnitudes for most of them. 
Here we examine the 2MASS, {\it HST}, and, where available, 
Johnson magnitudes from \citet{sho04} in order to estimate 
the reddening for each target.  Our approach is to calculate 
synthetic magnitudes in each band based on a consistent 
spectral flux distribution for a given ZAMS stellar effective 
temperature, $T_{\rm eff}$, and a given reddening, $E(B-V)$.  

We assembled a small grid of theoretical spectra covering the
wavelength range of interest for ZAMS stars along the upper main
sequence.  We used model spectra based on non-LTE, line-blanketed
model atmospheres from \citet{lan03} for the O-stars (for effective
temperatures of 27500, 35000, and 45000~K), and we adopted spectra
from LTE models by R.\ L.\
Kurucz\footnote{\url{http://kurucz.harvard.edu/grids/gridP00/fp00k4.pck}}
for the B-stars (for 10000, 15000, 20000, and 24000~K).  The gravity
$\log g$ for each temperature was taken from the 1~Myr ZAMS sequence
of \citet{lej01}.  In addition to the computational differences
between the Kurucz and Lanz \& Hubeny models, the former assume a
microturbulence of 4~km~s$^{-1}$ while the latter use 10~km~s$^{-1}$.
Despite these differences, the predicted flux distributions generally
make a smooth transition at the temperature boundary between them
(where non-LTE effects for hydrogen become negligible).

We calculated synthetic magnitudes for these spectra using 
the SYNPHOT routine CALCPHOT \citep{bus98}, 
which is distributed with the STSDAS package in IRAF. 
The software includes the spectral response functions for 
the {\it HST} wide band filters as well as those for 
the standard Johnson filters.   However, we used the 
response functions from \citet*{coh03} to simulate 
the sampling of the spectra in the IR with the 2MASS 
filter system.  The procedure returns synthetic magnitudes 
on the VEGAMAG system, so defined that Vega has a magnitude 0.0
in all filters.  Since Vega actually has magnitudes 
$U=0.02$, $B=0.02$, and $V=0.03$ in the Johnson system, 
we adjusted the both the Johnson and corresponding {\it HST} 
synthetic magnitudes by these amounts to make them 
consistent with observations.   The resulting synthetic 
ZAMS colors relative to Johnson $V$ and the 
absolute $K$ magnitude are collected in Table~3 
for our grid of stellar effective temperatures.  
These are generally in reasonable agreement with Johnson 
color indices published by \citet{weg94} and \citet*{bes98}
except perhaps for the shortest wavelengths and hottest stars. For example, \citet{tur76} and \citet{und79} argue that the hottest O-stars have $B-V = -0.32$ versus our lower limit of $B-V=-0.28$. 
Note that the $U-V$ and $B-V$ color indices are often
based on uncertain de-reddening procedures for O-type stars, 
which may tend to produce colors that are too blue. 
The lowest $B-V=-0.28$ for the 45000~K model is consistent
with that for the bluest O-stars in the Galaxy \citep{mai04}
and in the LMC \citep{fit88}.  

\placetable{tab3}      

We next transformed each of these model spectra for a grid 
of 40 values of $E(B-V)$ from 0 to 3.9 using the reddening 
scheme of \citet{fit99} for a ratio of total-to-selective 
extinction of $R=3.1$ (the same as adopted by \citealt{sho04}). 
We then used CALCPHOT to determine the predicted color indices as 
a function of the reddening.  A set of sample reddening curves 
is shown in Figure~2 for the 35000~K model spectrum.  
Note that unlike the other filters the color index trend for the 
{\it HST} F336W filter undergoes a reversal at large reddening 
because of a red-leak in this filter. 

\placefigure{fig2}     

We performed a numerical grid search to find the stellar temperature and 
reddening that best matched a given set of IR and optical color indices. 
We found this scheme was able to confirm the known reddening in a number 
of test cases for OB stars where Johnson $U$, $B$, and $V$ and 2MASS $J$, $H$, and $K$ 
magnitudes were available, but the method is not applicable to stars with temperatures cooler than our grid limit of 10000~K.  We applied the method to all the stars 
in our WFPC2 field for which 2MASS photometric data exist.  These stars 
are listed in Table~4, which gives our assigned number, the 2MASS designation, 
the number adopted by \citet{sho04}, our derived reddening, and de-reddened 
estimates of $m(336)-K$ and $K$ (see below).  There are several stars with 
relatively low reddening that are probably foreground objects, but most 
of the other stars have a reddening in the range $E(B-V)=1.5$ to 2.0 
(with the possible exception of the faint star \#8).  There are four 
cluster stars with a complete set of WFPC2, Johnson \citep{sho04},
and 2MASS magnitudes (after omission of the two WR stars for which the 
method may be poor), and the mean reddening for these four stars 
is $E(B-V) = 1.63\pm 0.05$ (mean error), in good agreement with the result of 
$E(B-V) = 1.60\pm 0.02$ from \citet{sho04}. 

\placetable{tab4}      

We can use these reddening fits to return to the question of magnitude
zero-point offsets that we raised in the previous section.  Table~5
lists the average values of the residuals from the fit for each filter
based on the same sample of four stars with complete magnitude
coverage.  These represent any systematical offsets between the
observed magnitudes and the synthetic magnitudes derived from the
theoretical spectra, so they encompass any lingering problems in the
observational zero-points, filter responses, and wavelength specific
deficiencies in the theoretical models.  The offsets in Table~5 are
generally small and comparable to the scatter in the residuals, but
there are some interesting exceptions.  We see, for example, that the
mean difference in the residuals for $m(555)-V$ is $0.21\pm0.04$,
which is the same within errors as the difference between the
transformed $V_{\rm HST}$ and $V_{\rm STP}$ that we found in the last
section.  Similarly the residuals in $m(439)-m(555)$ suggest that our
uncorrected color index is too blue in the same way as demonstrated in
our comparison of our transformed $B-V$ with the colors from
\citet{sho04}.  Thus, the mean residuals in Table~5 lead us to revise our original WFPC2 zeropoints to 18.305 for the F336W filter, 20.036 for the F439W filter, and 21.496 for the F555W filter. An application of these zero-point corrections should provide consistent results
across for the entire set of photometric measurements. We give in columns 5 and 6 of Table~4 the individually de-reddened color index $m(336)-K$ (including a zero-point offset of $-0.35\pm0.09$ from the mean residuals in Table~5) and extinction corrected $K$ (using a
zero-point offset of $+0.15\pm0.07$ from Table~5) that we will use in
the next section to estimate the cluster distance.

\placetable{tab5}      

We caution that these results are sensitive to the adopted value of 
total-to-selective extinction $R$.  This parameter is often estimated by 
comparing the reddening $E(B-V)$ to the extinction in the IR, and 
we used equations A3, A4, and A5 from \citet{fit99} to estimate $R$ 
for the four stars with complete Johnson and 2MASS photometry. 
The mean derived value, $R=3.25 \pm 0.15$, is consistent within errors 
with our assumed value of $R=3.1$, but the application of reddening curves 
for $R=3.25$ leads to a mean reddening of $E(B-V)=1.51 \pm 0.06$ for the 
same sample of four stars using the scheme outlined above.  We discuss below 
how such a revision in the reddening law influences the distance estimate. 

\section{Distance to the Cluster}  

We can estimate the distance to the cluster surrounding WR~38 and WR~38a 
based upon the positions of the stars in a de-reddened color and 
extinction corrected magnitude diagram.  The spectral flux distributions
of massive stars appear similar to the Rayleigh-Jeans tail of a 
blackbody spectrum in the optical and near-IR spectral range, and we need 
spectral line diagnostics (primarily through spectral classification) 
in order to estimate the stellar effective temperature reliably 
(and thus establish the luminosity of the star relative to the main
sequence).   In the absence of spectral data, the best approach is 
to use a color index based upon the widest wavelength range possible 
as a temperature parameter.  We selected the $m(336)-K$ index for 
this purpose, which spans a range of nearly 3 magnitudes over the OB 
star temperature range (compared with a range of only $\approx 0.26$ mag in $B-V$; 
see Table~3).  We also decided to use $K$ for the magnitude ordinate in 
the color--magnitude diagram since it is least affected by uncertainties 
in extinction.  

The color--magnitude diagram for the WR~38 cluster in this magnitude system
is illustrated in Figure~3.  The $m(336)-K$ colors were de-reddened 
and the $K$ magnitudes extinction corrected using the individual $E(B-V)$ 
values determined for each star in the last section (Table~4).  In addition, 
the zero-point corrections found from the mean residuals to the reddening solutions
(Table~5) were applied to bring the observed magnitudes into the synthetic 
magnitude system needed to compare the observed and theoretical 
color--magnitude diagrams.  There are five stars that have similar positions in 
Figure~3 that probably correspond to the cluster main sequence.  
One bright star, \#7, is located above the main sequence, and it is either 
an evolved, luminous star within the cluster or a foreground object.  
The two WR stars are found well to the right of the main sequence, 
which is probably due to the strong emission lines found in the $K$-band
\citep*{fig97}.   

\placefigure{fig3}     

We adopted the 1~Myr ZAMS from the models of \citet{lej01} to determine 
the distance modulus from the difference of the observed and theoretical
absolute magnitudes.  We compared our synthetic color index $m(336)-K$ 
values from SYNPHOT/CALCPHOT for our temperature grid with the
corresponding values from the tables of \citet{lej01}, and we found 
a mean difference of $0.04\pm0.06$ mag.  We added this small offset to the 
model ZAMS color index from \citet{lej01} in order to intercompare the colors 
consistently in the synthetic system.   We then compared the observed and 
model colors for a grid of assumed distance modulus values to determine 
the best fit estimate.  The derived distance modulus is 
$5\log d - 5 = 15.0^{+1.5}_{-1.0}$ based on the $\chi^2$ residuals of the fit
(corresponding to a distance of $d = 10^{+12}_{-4}$~kpc). 
At this distance, the five main sequence stars in Figure~3 have 
colors and magnitudes of ZAMS stars in the temperature range 
33000 to 38000~K, which are characteristic of O-type stars.

Our distance modulus estimate agrees within errors with that derived by \citet{sho04}, 
$15.80\pm0.25$.  The small difference in distance modulus between 
these investigations is probably due to differences in the adopted ZAMS isochrone. 
The ZAMS relation adopted by \citet{sho04} (from \citealt{tur76}) is an empirical relation based upon a 
somewhat evolved population of stars, and it is approximately 0.8~mag brighter at $B-V=-0.30$ 
than the theoretical 1~Myr ZAMS that we adopted from \citet{lej01}. 
Note that \citet{sch95} verified that the Geneva models used by \citet{lej01}  
yield absolute magnitudes that agree with those determined for eclipsing 
binaries after a correction is made for evolution from the ZAMS. 
We also show in Figure~3 the isochrone for an age of 8~Myr, approximately 
equal to the maximum age of clusters containing WR stars \citep*{mas01}, 
and a comparison of the isochrones shows that we may be underestimating 
the distance modulus by $\approx 1$~mag if the cluster is signftpificantly older than assumed. 
Unfortunately, we do not know the age of the cluster, and consequently the amount 
of evolutionary brightening at the upper end of the main sequence is
unknown.  Deeper photometric observations could resolve this problem 
through ZAMS fitting of less massive, unevolved stars. 

If the ratio of total-to-selective extinction is revised upwards to $R=3.25$
(\S3), then the derived reddening is smaller and the intrinsic $m(336)-K$ colors
are larger than shown in Figure~3.  Such a revision would lead to a decrease 
in distance modulus to a value of 13.5~mag, slightly below the $1~\sigma$ error 
limit quoted above for the $R=3.1$ model fit.  Thus, given the uncertainties 
in the cluster age and the reddening law, the range in acceptable distance 
modulus value may be somewhat larger than given above.  
Taken at face value, our analysis does not 
improve upon the the distance estimate from \citet{sho04}, $14.5\pm 1.6$~kpc, 
but we caution that their error refers only to internal sources and at 
this stage it is prudent to acknowledge an error range that accounts for 
systematical errors resulting from assumptions about the cluster isochrone
and reddening law. 

Our line of sight towards WR~38 ($l = 290\fdg57$, $b = -0\fdg92$) 
passes twice through the Carina spiral arm (see Fig.~4 of 
\citealt{gra88}) and we can use information from 
CO radio emission maps to help place the giant molecular clouds 
and associated massive star forming regions along this 
line of sight.   \citet{gra88} and \citet*{dam01}
show that there is a large and rare hole in the near side of 
the Carina spiral arm in this direction that might allow us to 
see some very distant clusters.  This first cut through the arm occurs at 
distance of $\approx 2.4$~kpc, and \citet{sho04} discuss some of 
the sparsely populated field stars they find at this distance. 
The next molecular cloud down the line is found at a distance of 
6.8~kpc in the direction $l = 290\fdg2$, $b = -0\fdg2$ 
(with a radial velocity of $-1$ km~s$^{-1}$ and 
designated \#14 in Fig.~2 of \citealt{gra88}), which
is located about $1\fdg0$ away from the WR~38 cluster. 
This distance is consistent with the distance error range for
the WR~38 cluster and the other nearby WR stars.  
However, just beyond this at 7.9~kpc we encounter one of the 
largest molecular clouds in the Carina spiral arm 
(with a radial velocity of $+22$ km~s$^{-1}$ and given as \#13 in 
Fig.~2 of \citealt{gra88}) where our line of sight 
crosses the spiral arm a second time.  The WR~38 cluster lies  
only $0\fdg4$ away from the edge of this cloud.  There is very little 
CO emission in the $(l,v)$ diagram corresponding to distances beyond this,
so it is doubtful that there are favorable environments for 
massive star formation beyond a distance of 8~kpc in this direction
(see also Fig.~5 in \citealt{rus03}). 
However, \citet{mcc04} present deep radio maps of 21~cm emission 
from neutral hydrogen in this region that suggest that some weak 
emission is present from an extension of the Perseus arm 
(at a distance of $\approx 15~$~kpc and a radial velocity of $+50$ km~s$^{-1}$)
and a possible outer arm (at a distance of $\approx 21~$~kpc 
and a radial velocity of $+120$ km~s$^{-1}$).  
The outer disks of large spiral galaxies occasionally show clear
H~I spiral arms extending to radii in excess of 30~kpc, 
while the stellar arms stop at radii of about 10~kpc.  A classic 
example is M83, where the H~I spiral arms extend three times as far as
the stellar arms \citep{til93}. 
Note that there are very few WR stars known in the Milky Way Galaxy beyond 
the solar circle (see Fig.~6 in \citealt{van01}), where they would be easy to
discover if they existed there (due to generally low reddening).  Furthermore,
the number ratio of WR to O-type stars decreases dramatically in low 
metallicity environments such as the outer Galaxy, so even if there are fair
numbers of O stars in the outer Galaxy, there will be relatively many fewer
WR stars there.

\citet{geo00} discuss the stars that are associated 
with H~II regions in this general direction, and they find that 
the extinction ranges from $A_V \approx 1.8$ at 2.8~kpc 
to $A_V \approx 2.6$ at 4.2~kpc, and up to $A_V \approx 4$ at 8~kpc. 
Our derived extinction of $A_V = 5.0 \pm 0.2$ would tend to suggest 
that the WR~38 cluster also resides at a distance of $\approx 8$~kpc or larger. 
The 8~kpc distance would place the WR~38 cluster approximately midway between 
molecular cloud \#13 (and associated H~II complex $289.3-0.6$; 
\citealt{geo00}) and molecular cloud \#17 
(H~II complex $291.6-0.7$) which is associated with the 
starburst cluster NGC~3603 \citep{mof02}. 
Note that \citet{sun04} find a reddening of $E(B-V)=1.8$ in the 
outskirts of NGC~3603 that is comparable to that for the WR~38 cluster. 
\citet{geo00} identify an H~II region $290.487-0.814$ within 
the larger $289.3-0.6$ complex that has as exciting stars 
\#48 and \#52 from the list of \citet{wra76}.  The two stars
have a comparable extinction ($A_V\approx 3.7$) and are located 
only $\approx 0\fdg3$ away from the cluster,
so these stars and the H~II region could be related to the 
WR~38/WR~38a cluster.   

The solution to the distance problem for the WR~38/WR~38a cluster
will come when spectra can be obtained for the cluster stars. 
Classification dispersion spectra would help establish the 
temperatures and luminosities of the stars, and thus determine 
accurately the distance modulus of the cluster.  Furthermore, 
radial velocities from spectra would also allow us to 
distinguish between the differential Galactic rotation 
expected for the locations along the Carina and Perseus spiral arms. 
Such observations would clearly help us understand the history of 
massive star formation in this distant region of the Galaxy. 

\acknowledgments

We are grateful to David Turner for sharing his results in advance of 
publication and for helpful comments on an earlier version of this work. 
We also thank Thomas Dame for comments about the molecular 
emission in this region of the sky. 
Support for {\it HST} Proposal numbers GO-6365 and GO-7282 was provided by NASA through 
grants from the Space Telescope Science Institute, which is operated by the
Association of Universities for Research in Astronomy, Inc., under
NASA contract NAS5-26555.  DJW is also grateful for the
support and financial assistance of the Georgia Space Grant
Fellowship Program. This research was published while DJW held a National Research Council Associateship Award at the National Aeronautics and Space Administration's Goddard Space Flight Center. AFJM thanks NSERC (Canada) and FCAR (Quebec) for financial support.
This publication makes use of data products from the 
Two Micron All Sky Survey, which is a joint project of the
University of Massachusetts and the Infrared Processing and 
Analysis Center/California Institute of Technology, funded
by the National Aeronautics and Space Administration and the 
National Science Foundation.



\clearpage

\begin{figure}
\caption{The {\it HST}/PC image of WR~38 (\#1) and WR~38a (\#21) 
made with the F555W filter (placed according to J2000 coordinates).
The identification number above each star corresponds to the 
system used in Tables 1 and 2.}
\label{fig1}
\end{figure}


\begin{figure}
\caption{Reddening curves based on synthetic magnitudes from 
SYNPHOT/CALCPHOT for a spectrum from a model with 
$T_{\rm eff}=35000$~K and $\log g = 4.16$ \citep{lan03} 
and for a reddening law with $R=3.1$ \citep{fit99}.  
The top panel shows how the {\it HST} color indices $m(336)-V$,
$m(439)-V$, and $m(555)-V$ vary with reddening $E(B-V)$. 
The $m(336)-V$ trend reverses at large reddening due to the red 
leak in the the F336W filter.  The middle panel shows the 
variations for the Johnson $U-V$ and $B-V$ indices, while the 
lower panel shows the same for the IR indices, 
$J-V$, $H-V$, and $K-V$. }
\label{fig2}
\end{figure}


\begin{figure}
\caption{A color-magnitude diagram based upon the de-reddened
color index $(m(336)-K)_0$ and the absolute $K$ magnitude 
(for a distance modulus of $5\log d - 5 = 15.0$). 
The solid (dashed) line represents the 1~Myr (8~Myr) isochrone from \citet{lej01}
while the plus signs represent the magnitudes and colors of the  
probable cluster stars.  Errors in the color index are represented by the 
horizontal line through each symbol (errors in $K$ are comparable
to the vertical symbol size).  The two WR stars that appear in the 
right hand side of the diagram and the bright star (\#7) located 
at the top of the diagram were omitted in the main sequence 
fitting procedure.}
\label{fig3}
\end{figure}


\clearpage

\begin{deluxetable}{ccccc}
\tablewidth{0pc}
\tablecaption{Astrometric Data for the WR~38/WR~38a Cluster\tablenotemark{a}}
\tablehead{
\colhead{Field} & 
\colhead{$\alpha$} & 
\colhead{$\delta$} & 
\colhead{Sep.\tablenotemark{b}} & 
\colhead{P.A.\tablenotemark{b}} \\
\colhead{Number} & 
\colhead{(J2000)} & 
\colhead{(J2000)} & 
\colhead{(\arcsec)} & 
\colhead{($^\circ$)} }
\startdata
\phn1\dotfill & 11:05:46.62 & $-$61:13:48.8 &  \nodata &     \nodata \\
\phn2\dotfill & 11:05:46.49 & $-$61:13:50.2 & \phn1.63 &       215.2 \\
\phn3\dotfill & 11:05:46.30 & $-$61:13:45.9 & \phn3.75 &       322.1 \\
\phn4\dotfill & 11:05:46.74 & $-$61:13:43.3 & \phn5.57 & \phn\phn8.9 \\
\phn5\dotfill & 11:05:47.56 & $-$61:13:55.3 & \phn9.38 &       133.4 \\
\phn6\dotfill & 11:05:47.32 & $-$61:13:40.3 & \phn9.90 & \phn   30.7 \\
\phn7\dotfill & 11:05:47.36 & $-$61:13:40.1 &    10.28 & \phn   31.5 \\
\phn8\dotfill & 11:05:48.23 & $-$61:13:47.1 &    11.79 & \phn   81.7 \\
\phn9\dotfill & 11:05:46.81 & $-$61:14:00.0 &    11.87 &       173.2 \\
   10\dotfill & 11:05:48.24 & $-$61:13:52.8 &    12.34 &       108.8 \\
   11\dotfill & 11:05:46.68 & $-$61:13:35.1 &    13.69 & \phn\phn1.8 \\
   12\dotfill & 11:05:45.79 & $-$61:13:35.8 &    14.32 &       335.3 \\
   13\dotfill & 11:05:45.33 & $-$61:13:36.9 &    15.08 &       322.0 \\
   14\dotfill & 11:05:48.63 & $-$61:13:41.7 &    16.21 & \phn   63.8 \\
   15\dotfill & 11:05:48.87 & $-$61:13:46.8 &    16.34 & \phn   82.8 \\
   16\dotfill & 11:05:46.41 & $-$61:13:31.3 &    17.59 &       355.1 \\
   17\dotfill & 11:05:46.21 & $-$61:13:31.1 &    17.95 &       350.6 \\
   18\dotfill & 11:05:47.72 & $-$61:13:32.5 &    18.17 & \phn   26.0 \\
   19\dotfill & 11:05:46.78 & $-$61:13:30.3 &    18.58 & \phn\phn3.6 \\
   20\dotfill & 11:05:44.81 & $-$61:13:35.1 &    18.91 &       316.4 \\
   21\tablenotemark{c}\dotfill & 11:05:49.14 & $-$61:13:41.6 &    19.59 & \phn   68.4 \\
   22\dotfill & 11:05:45.16 & $-$61:13:32.1 &    19.76 &       327.7 \\
   23\dotfill & 11:05:46.24 & $-$61:13:28.1 &    20.93 &       352.5 \\
   24\dotfill & 11:05:44.49 & $-$61:13:34.1 &    21.31 &       313.8 \\
   25\dotfill & 11:05:49.76 & $-$61:13:48.5 &    22.65 & \phn   89.2 \\
\enddata
\tablenotetext{a}{To convert to the 2MASS frame, add 0.17 s to $\alpha$ and add $0\farcs1$ to $\delta$.}
\tablenotetext{b}{With respect to WR~38 = target \#1.}
\tablenotetext{c}{WR~38a.}
\label{tab1}
\end{deluxetable}

\clearpage

\begin{deluxetable}{cccccc}
\tablewidth{0pc}
\tabletypesize{\footnotesize}
\tablecaption{Uncorrected Photometric Data for the WR~38/WR~38a Cluster}
\tablehead{
\colhead{Number} & 
\colhead{$m_{\rm F555W}$} & 
\colhead{$m_{\rm F439W}-m_{\rm F555W}$} & 
\colhead{$m_{\rm F336W}-m_{\rm F439W}$} & 
\colhead{$V$} & 
\colhead{$B-V$} }
\startdata
\phn1\tablenotemark{a}\dotfill &    15.00$\pm$0.03 &     1.21$\pm$0.05 &  
   0.37$\pm$0.06 & 14.97$\pm$0.03 & 1.13$\pm$0.05 \\
\phn\phn$=$ STP40\dotfill & \nodata & \nodata & \nodata & 14.66 & $1.28$ \\
\phn2\dotfill &    16.47$\pm$0.05 &     1.21$\pm$0.13 &    $-$0.18$\pm$0.14 & 
   16.44$\pm$0.06 &     1.13$\pm$0.13 \\
\phn\phn$=$ STP119\dotfill & \nodata & \nodata & \nodata & 16.21 & $1.23$ \\
\phn3\dotfill &    16.19$\pm$0.05 &     1.09$\pm$0.10 &     0.02$\pm$0.11 & 
   16.16$\pm$0.05 &     1.03$\pm$0.10 \\
\phn\phn$=$ STP101\dotfill & \nodata & \nodata & \nodata & 15.90 & $1.27$ \\
\phn4\dotfill &    17.60$\pm$0.12 &     1.35$\pm$0.42 &  \nodata & 
   17.57$\pm$0.12 &     1.24$\pm$0.42 \\
\phn5\dotfill &    16.81$\pm$0.06 &     1.32$\pm$0.19 &     0.13$\pm$0.24 & 
   16.79$\pm$0.07 &     1.21$\pm$0.19 \\
\phn\phn$=$ STP138\dotfill & \nodata & \nodata & \nodata & 16.56 & $1.33$ \\
\phn6\dotfill &    17.39$\pm$0.12 &  \nodata &  \nodata & 
 \nodata &  \nodata \\
\phn7\dotfill &    13.27$\pm$0.03 &     1.41$\pm$0.04 &     0.20$\pm$0.04 & 
   13.25$\pm$0.03 &     1.29$\pm$0.05 \\
\phn8\dotfill &    18.79$\pm$0.39 &  \nodata &  \nodata & 
 \nodata &  \nodata \\
\phn9\dotfill &    18.70$\pm$0.40 &  \nodata &  \nodata & 
 \nodata &  \nodata \\
   10\dotfill &    15.83$\pm$0.04 &     1.21$\pm$0.09 &     0.09$\pm$0.10 & 
   15.80$\pm$0.04 &     1.13$\pm$0.09 \\
\phn\phn$=$ STP76\dotfill & \nodata & \nodata & \nodata & 15.51 & $1.28$ \\
   11\dotfill &    18.63$\pm$0.32 &  \nodata &  \nodata & 
 \nodata &  \nodata \\
   12\dotfill &    18.69$\pm$0.35 &  \nodata &  \nodata & 
 \nodata &  \nodata \\
   13\dotfill &    16.96$\pm$0.07 &     0.85$\pm$0.16 &     0.28$\pm$0.20 & 
   16.94$\pm$0.08 &     0.82$\pm$0.16 \\
   14\dotfill &    16.74$\pm$0.07 &     1.42$\pm$0.22 &     0.08$\pm$0.26 & 
   16.72$\pm$0.07 &     1.30$\pm$0.22 \\
\phn\phn$=$ STP137\dotfill & \nodata & \nodata & \nodata & 16.51 & $1.30$ \\
   15\dotfill &    17.55$\pm$0.13 &     1.41$\pm$0.43 &  \nodata & 
   17.53$\pm$0.13 &     1.29$\pm$0.43 \\
   16\dotfill &    17.11$\pm$0.09 &     1.24$\pm$0.25 &     0.00$\pm$0.29 & 
   17.09$\pm$0.09 &     1.15$\pm$0.26 \\
   17\dotfill &    17.83$\pm$0.16 &     1.22$\pm$0.47 &  \nodata & 
   17.80$\pm$0.16 &     1.13$\pm$0.47 \\
   18\dotfill &    18.79$\pm$0.40 &  \nodata &  \nodata & 
 \nodata &  \nodata \\
   19\dotfill &    16.00$\pm$0.04 &     1.07$\pm$0.09 &    $-$0.04$\pm$0.10 & 
   15.98$\pm$0.05 &     1.01$\pm$0.09 \\
   20\dotfill &    17.99$\pm$0.17 &     0.83$\pm$0.39 &  \nodata & 
   17.97$\pm$0.17 &     0.80$\pm$0.39 \\
   21\tablenotemark{b}\dotfill &    15.33$\pm$0.04 &     1.24$\pm$0.07 & 
   $-$0.13$\pm$0.08 & 15.30$\pm$0.04 & 1.15$\pm$0.07 \\
\phn\phn$=$ STP57\dotfill & \nodata & \nodata & \nodata & 15.12 & $1.15$ \\
   22\dotfill &    18.12$\pm$0.21 &  \nodata &  \nodata & 
 \nodata &  \nodata \\
   23\dotfill &    15.38$\pm$0.04 &     0.45$\pm$0.05 &     0.42$\pm$0.05 & 
   15.36$\pm$0.04 &     0.45$\pm$0.06 \\
   24\dotfill &    18.22$\pm$0.23 &  \nodata &  \nodata & 
 \nodata &  \nodata \\
   25\dotfill &    17.50$\pm$0.13 &     1.23$\pm$0.38 &  \nodata & 
   17.47$\pm$0.13 &     1.15$\pm$0.38 \\
\enddata
\tablenotetext{a}{WR~38.}
\tablenotetext{b}{WR~38a.}
\tablecomments{Corresponding stars and results from
 \citet{sho04} (STP\#) are listed below their {\it HST} values.}
\label{tab2}
\end{deluxetable}

\clearpage

\begin{deluxetable}{lccccccccc}
\tablewidth{0pc}
\tabletypesize{\footnotesize}
\tablecaption{ZAMS Intrinsic Colors from Synthetic Magnitudes}
\tablehead{
\colhead{$T_{\rm eff}$} & 
\colhead{$m(336)-V$} & 
\colhead{$m(439)-V$} & 
\colhead{$m(555)-V$} & 
\colhead{$U-V$} & 
\colhead{$B-V$} & 
\colhead{$J-V$} & 
\colhead{$H-V$} & 
\colhead{$K-V$} & 
\colhead{$K_0$} }
\startdata
 10000 &  $-$0.15 &  $-$0.02 &  $-$0.00 &  $-$0.07 &  $-$0.02 &   0.06 &   0.08 &   0.08 & \phs1.46 \\
 15000 &  $-$1.04 &  $-$0.14 &  $-$0.01 &  $-$0.68 &  $-$0.14 &   0.34 &   0.39 &   0.44 & \phs0.46 \\
 20000 &  $-$1.42 &  $-$0.20 &  $-$0.02 &  $-$0.95 &  $-$0.19 &   0.49 &   0.57 &   0.63 &  $-$0.44 \\
 24000 &  $-$1.63 &  $-$0.22 &  $-$0.02 &  $-$1.10 &  $-$0.22 &   0.58 &   0.67 &   0.76 &  $-$1.09 \\
 27500 &  $-$1.75 &  $-$0.26 &  $-$0.02 &  $-$1.20 &  $-$0.25 &   0.63 &   0.72 &   0.82 &  $-$1.66 \\
 35000 &  $-$1.90 &  $-$0.28 &  $-$0.02 &  $-$1.32 &  $-$0.27 &   0.70 &   0.81 &   0.92 &  $-$2.97 \\
 45000 &  $-$1.96 &  $-$0.28 &  $-$0.02 &  $-$1.36 &  $-$0.28 &   0.71 &   0.82 &   0.93 &  $-$4.79 \\
\enddata
\label{tab3}
\end{deluxetable}

\clearpage

\begin{deluxetable}{ccccccl}
\tablewidth{0pc}
\tabletypesize{\footnotesize}
\tablecaption{$R=3.1$ Reddening Parameters for Stars with 2MASS Photometry}
\tablehead{
\colhead{} & 
\colhead{} & 
\colhead{} & 
\colhead{} & 
\colhead{$(m(336)-K)$} & 
\colhead{} & 
\colhead{} \\
\colhead{Field \#} & 
\colhead{2MASS \#} & 
\colhead{STP \#} & 
\colhead{$E(B-V)$} & 
\colhead{$-E(m(336)-K)$} & 
\colhead{$K-A(K)$} & 
\colhead{Comment} }
\startdata
\phn1 & 11054642$-$6113487 & \phn40  & $1.46\pm0.12$ & $-1.35\pm0.06$   &    $10.43\pm0.05$ & WR 38 \\
\phn3 & 11054613$-$6113457 &    101  & $1.51\pm0.09$ & $-2.68\pm0.10$   &    $12.16\pm0.07$ & all bands \\
\phn5 & 11054738$-$6113550 &    138  & $1.70\pm0.09$ & $-2.81\pm0.17$   &    $12.31\pm0.03$ & all bands \\
\phn7 & 11054719$-$6113398 & \nodata & $1.84\pm0.11$ & $-2.91\pm0.03$   & $\phn8.36\pm0.02$ & \nodata \\
\phn8 & 11054805$-$6113469 & \nodata & $2.51\pm0.24$ & \nodata          &    $11.97\pm0.06$ & \nodata \\
\phn9 & 11054664$-$6114003 & \nodata & $1.70\pm0.20$ & \nodata          &    $14.24\pm0.12$ & \nodata \\
   10 & 11054806$-$6113526 & \phn76  & $1.61\pm0.10$ & $-2.71\pm0.09$   &    $11.54\pm0.06$ & all bands \\
   12 & 11054562$-$6113358 & \nodata & $0.20\pm0.10$ & \nodata          & \nodata           & foreground \\
   13 & 11054515$-$6113367 & \nodata & $0.70\pm0.08$ & \nodata          & \nodata           & foreground \\
   14 & 11054845$-$6113414 &    137  & $1.69\pm0.15$ & $-2.62\pm0.17$   &    $12.16\pm0.08$ & all bands less $J$\\
   15 & 11054869$-$6113465 & \nodata & $1.47\pm0.06$ & \nodata          &    $13.21\pm0.09$ & \nodata \\
   19 & 11054659$-$6113301 & \nodata & $1.52\pm0.10$ & $-2.98\pm0.09$   &    $12.13\pm0.06$ & \nodata \\
   21 & 11054896$-$6113413 & \phn57  & $1.49\pm0.09$ & $-1.60\pm0.06$   &    $10.33\pm0.04$ & WR 38a \\ 
   23 & 11054607$-$6113279 & \nodata & $0.55\pm0.04$ & $-0.56\pm0.12$   &    $13.84\pm0.11$ & foreground \\
   25 & 11054956$-$6113482 & \nodata & $1.39\pm0.09$ & \nodata          &    $13.28\pm0.06$ & \nodata \\
\enddata
\label{tab4}
\end{deluxetable}

\clearpage

\begin{deluxetable}{lc}
\tablewidth{0pc}
\tablecaption{Average Magnitude Residuals from Reddening Fits}
\tablehead{
\colhead{Filter} & 
\colhead{$<m_\lambda({\rm observed})-m_\lambda({\rm synthetic})>$} }
\startdata
F336W &\phs$0.20\pm0.05$ \\
F439W &\phs$0.04\pm0.04$ \\
F555W &\phs$0.23\pm0.04$ \\
$U$   &$-0.17\pm0.06$    \\
$B$   &\phs$0.03\pm0.06$ \\
$V$   &\phs$0.02\pm0.01$ \\
$J$   &\phs$0.00\pm0.01$ \\
$H$   &$-0.06\pm0.09$    \\
$K$   &$-0.15\pm0.07$    \\
\enddata
\label{tab5}
\end{deluxetable}


\clearpage
\setcounter{figure}{0}
\begin{figure}
\plotone{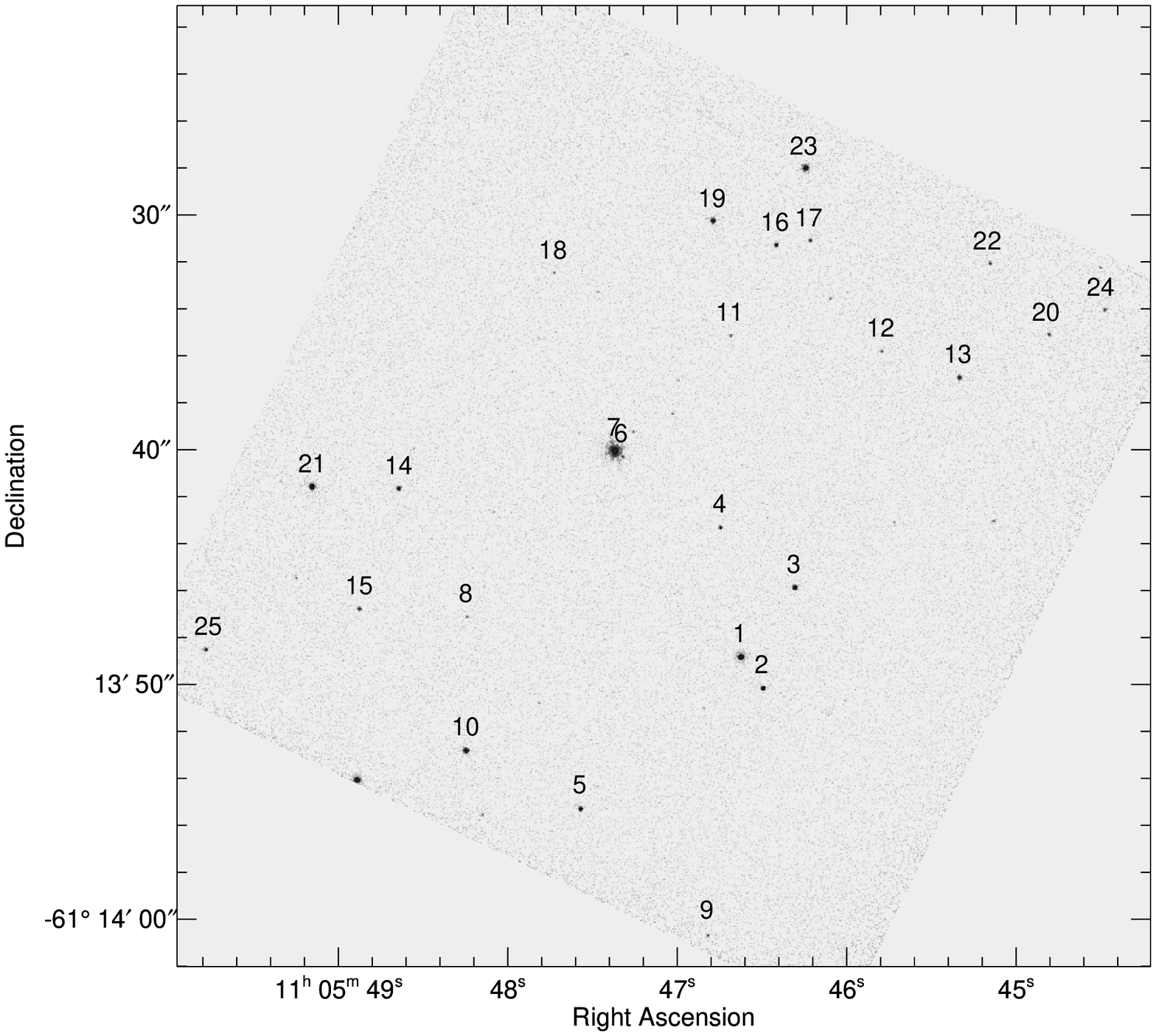}
\caption{}
\end{figure}

\clearpage

\begin{figure}
\plotone{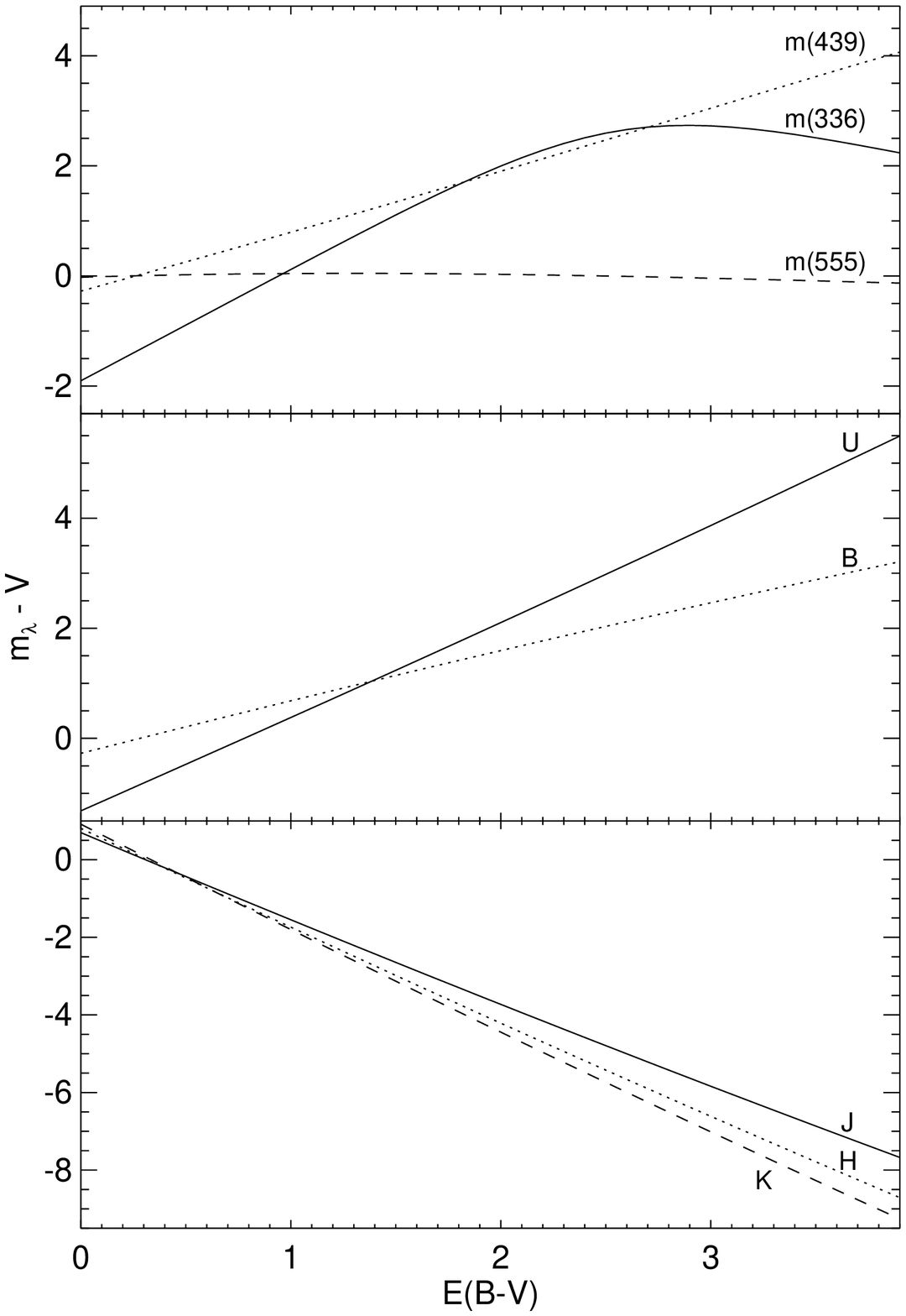}
\caption{}
\end{figure}

\clearpage 

\begin{figure}
\plotone{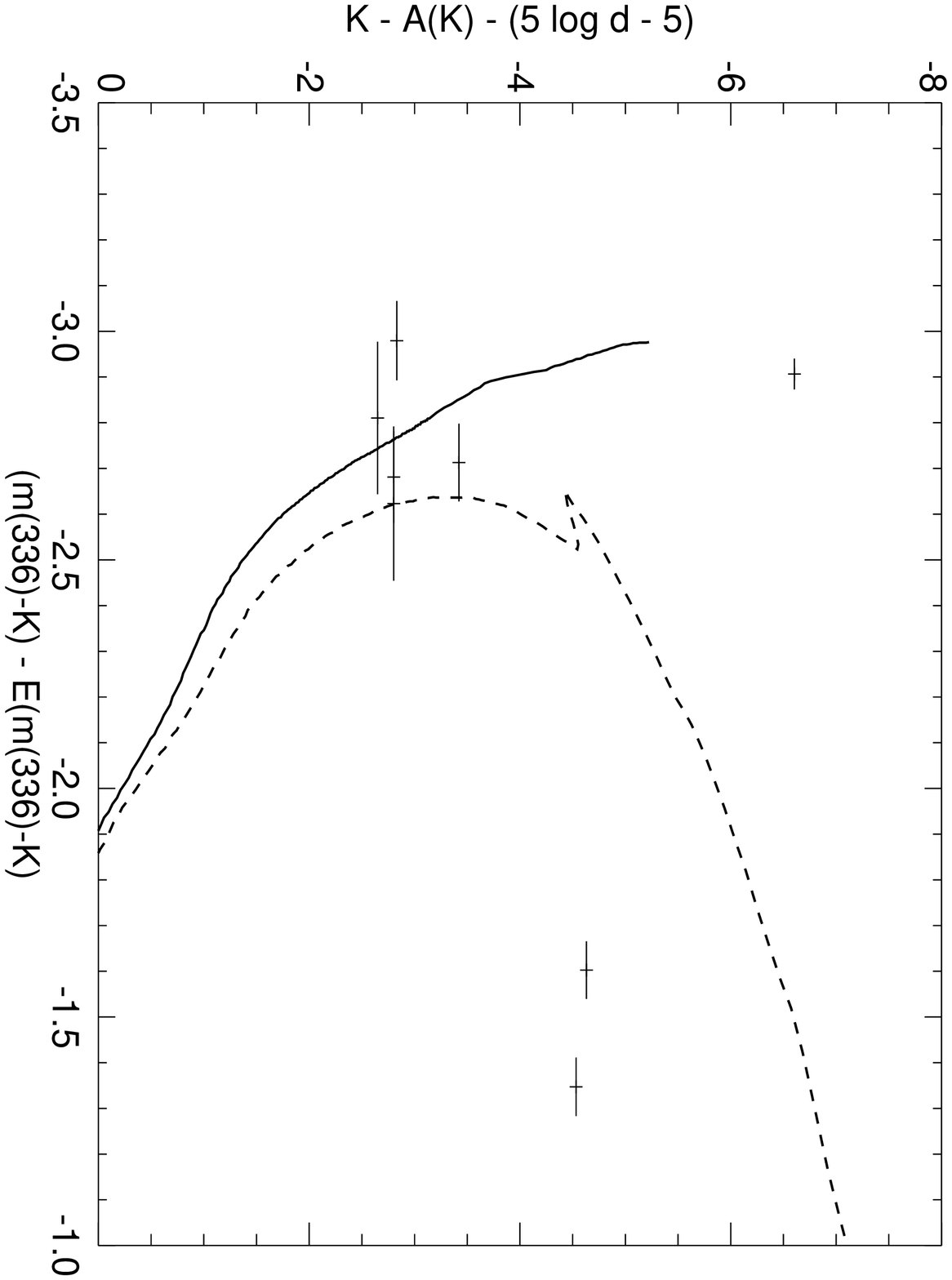}
\caption{}
\end{figure}

\end{document}